\newcommand{\di}{\mathrm{d}}
\documentclass[11pt,a4paper,twoside]{report}
\usepackage{mathtools}
\usepackage[utf8]{inputenc}
\usepackage[T1]{fontenc}
\usepackage[english]{babel}
\usepackage{amsmath}
\usepackage{multicol}
\usepackage{amssymb}
\usepackage{hyperref}
\usepackage{float}
\usepackage{caption}
\usepackage{tabularx}
\usepackage{booktabs}
\usepackage{longtable}
\usepackage[top=3cm,bottom=3cm,left=3cm,right=3cm]{geometry}
	
\begin{document}
\setlength\parindent{10pt} 	
	\newcommand{\thetitle}{ASPECTS OF NEUTRINO INTERACTIONS \\ (Scatterings at the small $Q^2$-Region)}
	\thispagestyle{empty}
	\begin{center}
		\begin{flushright}
			DO-TH 14/31
		\end{flushright}
		\vspace{1cm}
		\textbf{\Large{\thetitle}}\\
		\vspace{1cm}
		\small T. Hoinka\footnote{Email Address: \textbf{\texttt{tobias.hoinka@tu-dortmund.de}}}, \ \small E. A. Paschos\footnote{Email Address: \textbf{\texttt{paschos@physik.uni-dortmund.de}}},\ \small L. Thomas\footnote{Email Address: \textbf{\texttt{leo.thomas@tu-dortmund.de}}} \\
		\ \small \textit{Department of Physics, Technical University of Dortmund, D-44221, Dortmund, Germany} \\
		\vspace{.5cm}
		(Presented by E. A. Paschos at the CETUP-Workshop on Neutrino Interactions, July 22-31, 2014 at Lead/Dead Wood, South Dakota, USA)
		\vfill
		\begin{minipage}{13cm}
			\textbf{Abstract:} The article begins with a description of chiral symmetry and its application to neutrino induced reactions. For small $Q^2$ (forward direction) the process is dominated by the amplitute with helicity zero where the pion pole disappears when multiplied with the polarization vector. The remaining part of the amplitude is determined by PCAC. For $E_\nu > 2$\,GeV the computed cross sections are in good agreement with data. In coherent pion production we expect equal yields for neutrinos and antineutrinos a relation which for $E_\nu > 2$\,GeV is fulfilled. We discuss specific features of the data and suggest methods for improving them by presenting new estimates for the incoherent background. 
		\end{minipage}
		\vspace{2pt}
	\end{center}
\newpage
\vspace{.8cm}\noindent\textbf{\Large{1. INTRODUCTION}}\vspace{.8cm}\\

In neutrino reactions the small $Q^2$ region provides the opportunity for estimating the vector and axial contributions accurately. We will concentrate in this region and describe the methods in some detail. We will discuss two processes:
\begin{enumerate}
	\item[i.] Coherent pion production on Nuclei, and
	\item[ii.] The production of the Delta resonance and its subsequent decay into a pion and a nucleon, where the propagation and development of the final state is influenced by nuclear corrections.
\end{enumerate}
There are two schools for pion production calculations at low $Q^2$: the traditional based on PCAC (partially conserved axial current hypothesis) [1-5] and microscopic calculations [6-11]. The methods were reviewed in a recent article [12]. In estimates of PCAC there is the question how large is the region of validity and will be addressed later in the article.

\begin{figure}[H]
\begin{center}
\begin{minipage}{.45\linewidth}
	\centering
	\includegraphics[width=.9\linewidth]{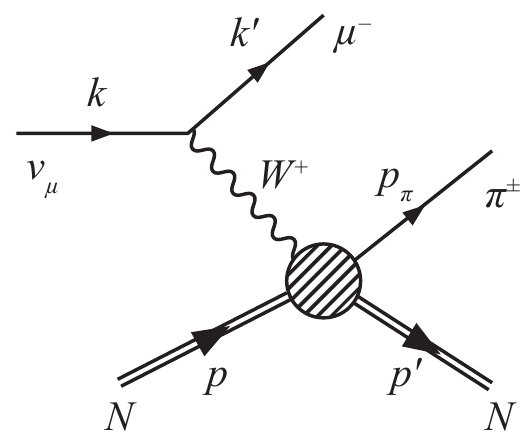}
\end{minipage}
\begin{minipage}{.45\linewidth}
	\centering
	\includegraphics[width=.8\linewidth]{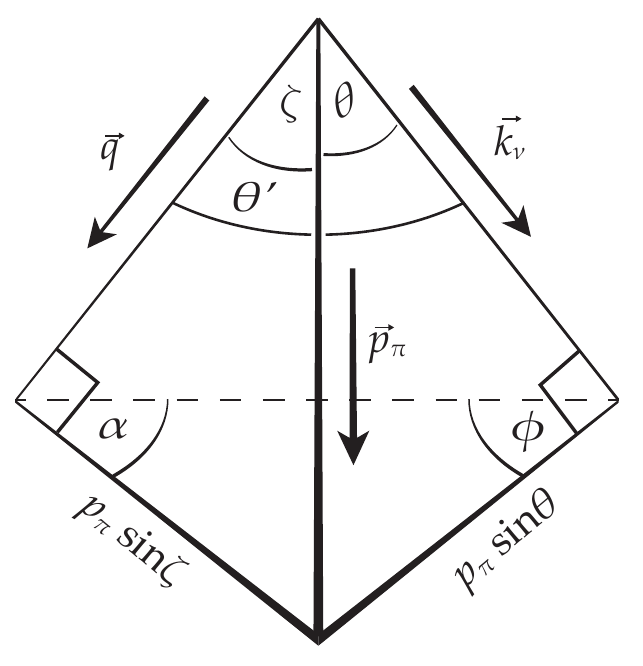}
\end{minipage}
\caption{Left: Feynman diagram for the charged current reaction. Right: Coordinate system used for angular dependance.}\label{fig1}
\end{center}
\end{figure}

In the region we consider the current from the leptonic vertex is expanded in terms of four polarization vectors. Three of them have helicity one and the scalar has helicity zero. The geometry of the process is defined in figure (1) where the weak current (virtual W or Z) moves along the z-axis and together with the neutrino direction they define the x-z-plane. The pion can be produced outside this plane. The three vectors of momenta form a tetrahedron. The four-vector of the virtual current has the components
\begin{align*}
	q_\mu =(q_0, 0, 0, q_z)
\end{align*}
and the polarization vectors with angular momentum one are 
\begin{align*}
	\epsilon_\mu (\lambda=\pm1) = \frac12\begin{pmatrix}0\\1\\\pm i\\0\end{pmatrix},\qquad\epsilon_\mu(\lambda=0) = \frac{1}{\sqrt{Q^2}}\begin{pmatrix}\vert\vec q\vert\\0\\0\\\nu\end{pmatrix}
\end{align*}
In addition there is the scalar polarization vector
\begin{align}
	\epsilon_\mu(l) = \frac{q_\mu}{\sqrt{Q^2}}.\label{eq1}
\end{align}
For small $Q^2$ there is a property of the weak interactions that simplifies the calculations. For $\nu^2 \gg Q^2$ the leptonic current for charged and neutral reactions is dominated by the polarizations $\epsilon_\mu(\lambda=0)$ and $\epsilon_\mu(l)$. In general, the amplitude for the axial hadronic current is written as the sum of a pion pole and a smooth remainder
\begin{align}
	-i\left<\pi^+N\right|\mathcal{A}_\mu^+\left|N\right> = \frac{\sqrt{2}f_\pi q_\mu}{Q^2+m_\pi^2}T(\pi^+N\to\pi^+ N)-\mathcal{R}_\mu,\label{eq2}
\end{align}
where $T(\pi^+N\to\pi^+ N)$ is $\pi+N$ elastic amplitude. The divergence of the matrix element on the left hand side of the equation is determined by PCAC. When this result is combined with the pion pole on the right-hand side the pion propagator cancels and the remainder gives a smooth function for $q_\mu \mathcal{R}_\mu$
\begin{align}
	q^\mu \mathcal{R}_\mu = -\sqrt{2}f_\pi T(\pi^+N\to\pi^+ N)\label{eq3}
\end{align}
In reference [3] the transverse contribution to coherent scattering was estimated and was found that it is small relative to the contribution for zero helicity. The important polarizations produce the cross section [3,4]
\begin{align}
	\frac{\mathrm{d}\sigma}{\mathrm{d}Q^2\mathrm{d}\nu\mathrm{d}t} = \frac{G_F^2\vert V_\mathrm{ud}\vert^2}{8\pi^2}\frac{\nu f_\pi^2}{E_\nu^2 Q^2}\left\{\tilde L_{00} + 2\tilde L_{l0}\frac{m_\pi^2}{Q^2+m_\pi^2}+\tilde L_{ll}\left(\frac{m_\pi^2}{Q^2+m_\pi^2}\right)^2\right\}\frac{\mathrm{d}\sigma_\pi}{\mathrm{d}t}\label{eq4}
\end{align}

The notation in this equation is standard with $E_\nu$, $\nu$ and $Q^2$ defined to be the energy of the neutrino, the energy transfer between the leptons and $q^2= - Q^2$ the square of the four momentum of the current, respectively. The pion decay coupling constant $f_\pi = 0.093\,$GeV and $L_{00}$, $L_{l0}$ and $L_{ll}$ are density matrix elements arising from the polarizations of the leptonic tensor [3,4]. We shall use equation (\ref{eq4}) for coherent scattering and for the axial contribution to the excitation of resonances. 

Coherent scattering is defined as scattering on a nucleus which remains in its ground state. Thus there is no exchange of quantum numbers between the virtual current-pion system and the nucleus and only the axial current contributes, especially in the two polarizations included in equation (\ref{eq4}). For coherent scattering the cross section $\di\sigma(\pi N)/\di t$ denotes the elastic scattering of a pion on the nucleus N. The size of the nucleus is large relative to that of protons and the form factor obtained as a Fourier transform of the nuclear density is a fast falling function of $t =(q- p_\pi)^2$. Data for elastic pion-Carbon scattering are available [13-15] and will be used in section 3. For the background we take the incoherent sum of pion-proton and pion-neutron scattering and fold them with nuclear corrections (final state interactions).
The amplitudes computed with the PCAC relation are propotional to the square of the pion mass and go to zero as $m_\mu\to0$. In our case the divergence of the hardonic matrix element has this property and it cancels the pion pole leaving a smooth function. In other words chiral symmetry determines a term which cancels the pion pole with the remainder being a smooth function. The remainder is transformed into a physical process for which we shall use experimental data. Using pion-nucleus scattering data includes an implicit assumption that the amplitudes do not change much when the pion is taken off the-mass-shell. This is correct for small values of $Q^2$ and we must face the question how large can $Q^2_\mathrm{max}$ be in each process.

\vspace{.8cm}\noindent\textbf{\Large{2. KINEMATICS}}\vspace{.8cm}\\
For neutrino induced reactions the calculation must respect the physical boundaries dictated by the kinematics of the lepton vertex. In particular, the four momentum of the current is space-like so that the variable $t = (q-p_\pi)^2$ can not reach the value $t = 0.0$. Any integration over $\vert t\vert$ covers the range
\begin{align}
	\left(\frac{Q^2+m_\pi^2}{2\nu}\right)^2 < \vert t\vert < \vert t\vert_0\label{eq5}
\end{align}

The upper limit of integration is the first diffractive minimum and can be extended to infinity without any noticeable change in the numerical results. The lower limit for the value of $\vert t\vert$ is important because $\frac{\di\sigma}{\di t}(\pi A)$ is a very fast falling function of $\vert t\vert$ with the following consequence: when one integrates to $\vert t\vert\to0$ there is a sizable overestimate of the neutrino induced cross section (by almost a factor of two) [4].

The calculation of the incoherent background is more complicated and has different kinematics. It receives a contribution from the vector current and the vector\,$\otimes$\,axial interference term. For the kinematic variables we use the notation from figure (\ref{fig1}). From the square of the vector $p^\prime = (q-p_\pi )+p$ and the definition of $t$ we obtain
\begin{align*}
	\nu = E_\pi-\frac{t}{2M},
\end{align*}
which in the approximation $\nu E_\pi \sim \vert\vec q\vert\vert\vec p_\pi \vert$ gives
\begin{align}
	\nu \approx \frac{ME_\pi+\frac12\left[Q^2-m_\pi^2\right]}{M-E_\pi\left(1-\cos\zeta\right)}\label{eq6}
\end{align}

The mass $M$ is the mass of the target, which for a nucleus is larger than the other quantities . This gives the relation $\nu \sim E_\pi$ that we frequently use for coherent scattering.

Balancing the three momenta of the muon and the current perpendicular to the neutrino direction gives
\begin{align}
	Q^2 \approx \frac{2E_\nu E_\pi^2}{E_\nu\cos\theta^\prime-E_\pi}(1-\cos\theta^\prime)\label{eq7}
\end{align}
Two more relations follow from the geometry of the tertahedron in figure (\ref{fig1})
\begin{align}
	\sin\alpha = \frac{\sin\theta}{\sin\zeta}\sin\phi\label{eq8}
\end{align}
and the addition theorem 
\begin{align}
	\cos\zeta = \cos\theta\cos\theta^\prime+\sin\theta\sin\theta^\prime\cos\phi\label{eq9}
\end{align}
These relations are useful when we wish to change the variables referring to the current to those referring to the neutrino direction.

\newpage
\vspace{.8cm}\noindent\textbf{\Large{3. THE CROSS SECTIONS}}\vspace{.8cm}\\
The most detailed evidence for coherent scattering is the observation of the characteristic sharp peak in the t distribution of the events. This was the evidence in the early experiments. In the analysis of the Minerva data the background is estimated by studying the $t$-distribution of pions for $0.2<\vert t\vert <0.6\,$Gev$^2$ and then extrapolating the functional form to smaller values, where the signal for coherent scattering peaks at $\vert t\vert$ near zero. Distributions on the variable $\vert t\vert $ are not available but integrals over $\vert t\vert $ for various neutrino energies were presented at this meeting [16]. Instead of calculating the cross section again, we take the figures presented some time ago in reference [4] and plot in the figures the new experimental points.

The elastic pion-Carbon cross section was parametrized as follows:
\begin{align}
	\frac{\di\sigma(\pi N)}{\di t} = a\exp\big[-b\vert t\vert\big]\label{eq10}
\end{align}
with values for the paramaters given in table (1) of [4] . The limits of integration were described in the previous section. The values of $a(\nu)$ and $b(\nu)$ are given in the table and go up $1.046$ GeV. Beyond that value the data were extrapolated to have constant values equal to the last entry in the table, i.e. $a = 3.53$ (barn/GeV$^2$) and $b=53.49$ (1/GeV$^2$). As we mentioned in the introduction, only the axial current contributes to coherent scattering, expecting neutrino and antineutrino induced reactions to be equal. The data is consistent with this property and we shall use the same curves for both reactions. In article [4] the cross section was calculated up to $E_\nu=10\,$GeV. A similar calculation appeared in reference [5] for a smaller range of $E_\nu$ up to $2.0\,$GeV; up to this energy the results of the two groups agree.

\begin{figure}[htpb]
	\begin{center}
	\includegraphics[width=12cm]{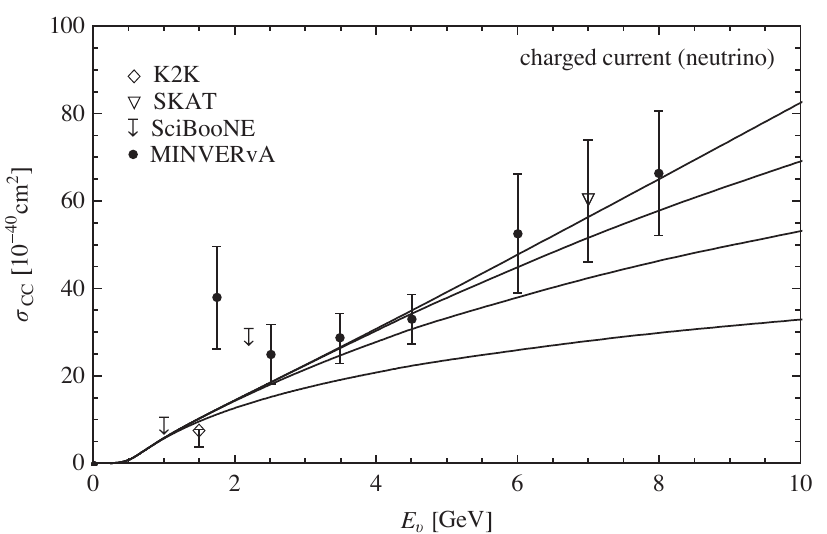}
	\caption{Integrated charged current cross section wit $Q_\mathrm{max}^2 = 0.2,0.5,1.0$ and $4.4\,$GeV$^2$ (bottom to top). The curves are from [4] and the data from [16,17,18,19].}
	\label{fig2}
	\end{center}
\end{figure}

\begin{figure}[htpb]
	\begin{center}
	\includegraphics[width=12cm]{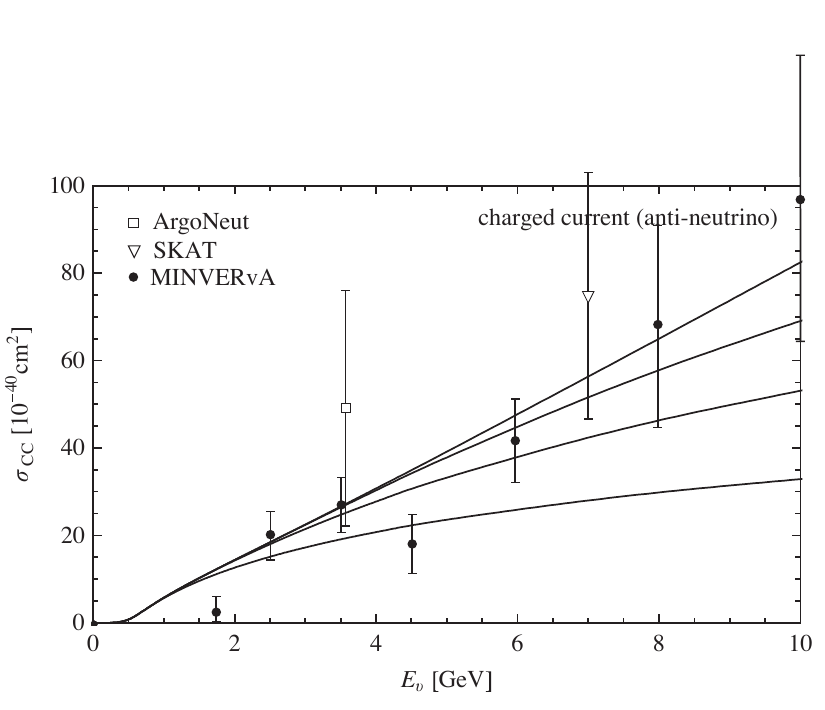}
	\caption{Integrated charged current cross section. The curves are the same as in figure \ref{fig2} and the data for antineutrinos from ArgoNeut [20] and MINERvA [16] and SKAT [19]}
	\label{fig3}
	\end{center}
\end{figure}

In figure (\ref{fig2}) we show the neutrino data. We include the older data from K2K [17], SciBoone [18] and SKAT [19] and to them we added the results from Minerva reported at the Workshop [16, Higuera]. A point from ArgNeut has a very large error and lies outside the figure. In figure (\ref{fig3}) we present the same curves and added the experimental points for antineutrinos from ArgoNeut [20] and Minerva. For $E_\nu>2.0\,$GeV the agreement between theory and experiment is very good, however the errors for antineutrinos are larger. In these energies neutrino and antineutrino cross sections are consistent to being equal. Only at the first point with an average energy between $1.5$ and $2.0\,$GeV the values of the two cross sections are different. These points do not seem to follow the trend of the higher points or of the curves and we discuss them below.

\begin{figure}[htbp]
	\begin{center}
	\includegraphics[width=11cm]{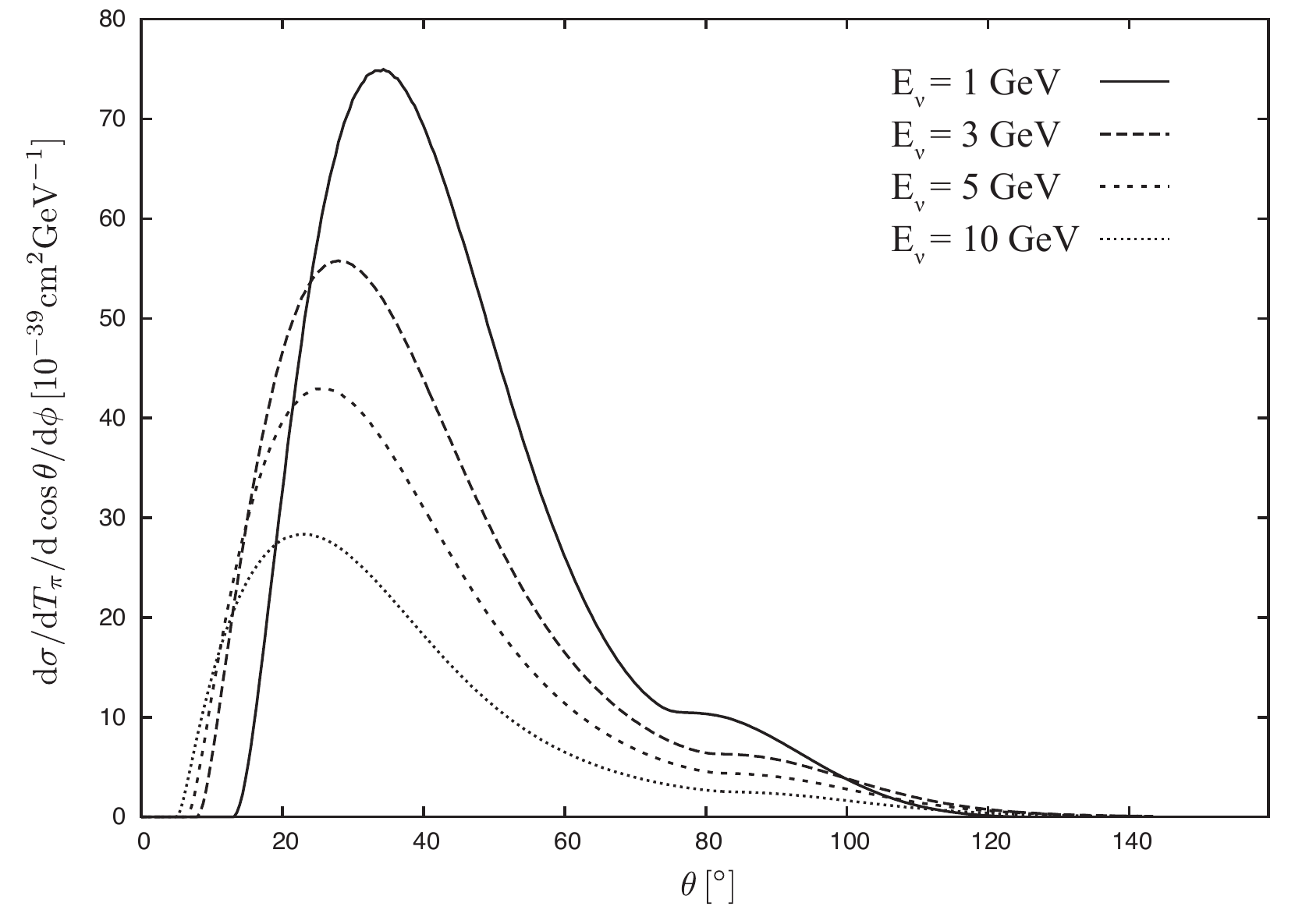}
	\caption{Distributions on the polar angle for a $^{12}$C target and for various values of $E_\nu$, for $E_\pi = 0.4\,$GeV and $\phi = 0$}
	\label{fig4}
	\end{center}
\end{figure}

\begin{figure}[htbp]
	\begin{center}
	\includegraphics[width=11cm]{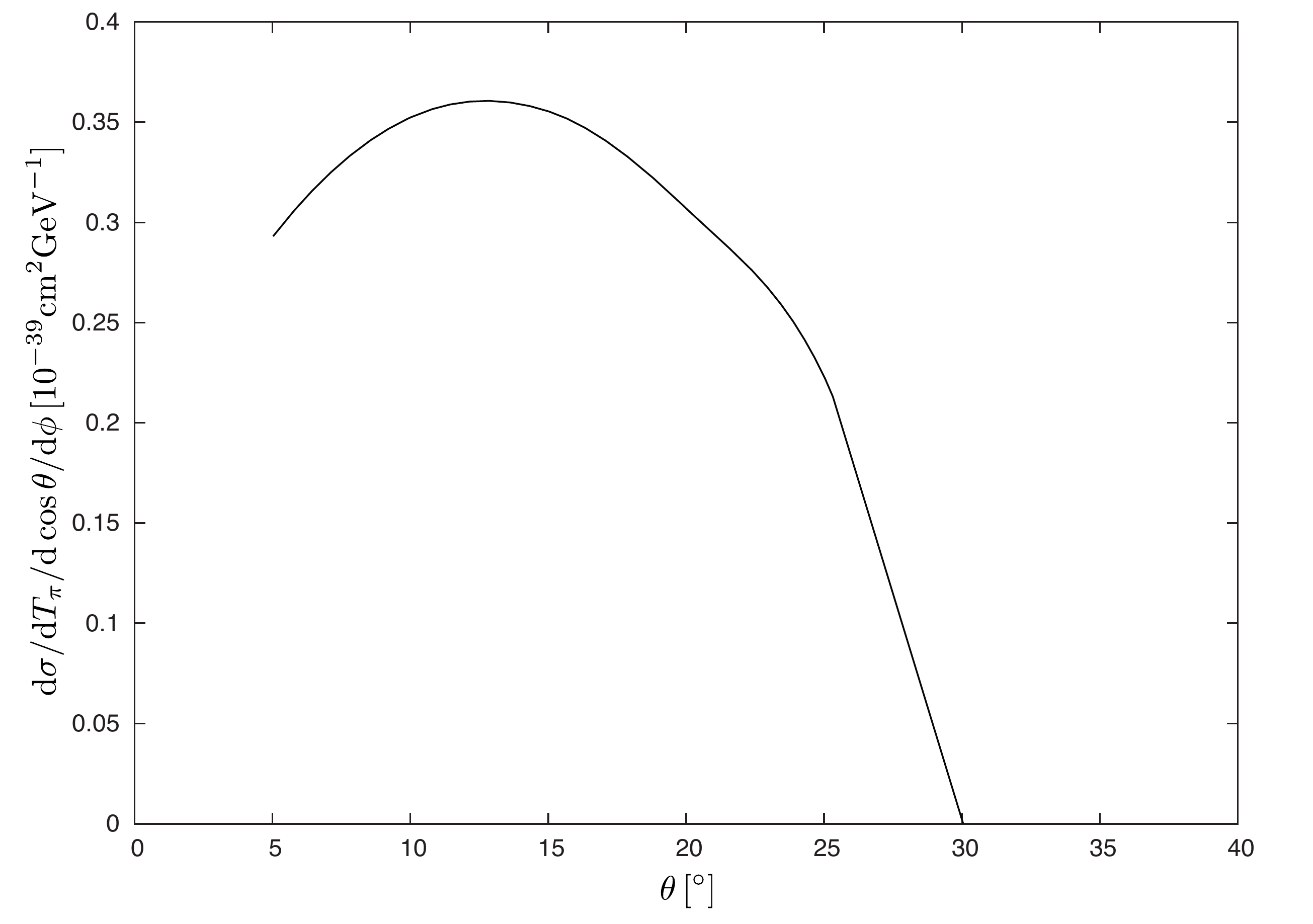}
	\caption{Triple differential cross section for coherent scattering for $E_\nu=1.0\,$GeV, $E_\pi=0.4\,$GeV and $\phi=0$}
	\label{fig5}
	\end{center}
\end{figure}

The PCAC approach is valid for low values of $Q^2$. In the comparison with the curves, the experimental points for energies up $4\,$GeV agree with the curves with the $Q^2$ having small values, below $0.5\,$Gev$^2$. For the higher energies the points agree with curves whose $Q^2_\mathrm{max}$ are closer to $1.0\,$GeV. This is a consequence of the fact that we used for $1.064<\nu<10\,$GeV constant values for $a$ and $b$ occuring in equation (\ref{eq8}). If we use for $b$ values smaller than $53.49$ (1/GeV$^2$), then similar curves are obtained for smaller values of $Q^2$. In addition, at this low energy the background is dominated by the production of the $\Delta$-resonance whose angular distribution is $(1+3\cos^2\theta^\ast)$ where $\theta^\ast$ and all other starred quantities refer to center of mass. The contribution of the axial current to the cross section in the cms system is
\begin{align}
	\frac{\di\sigma}{\di Q^2 \di\nu\di\cos\theta^\ast} = \frac{G_F^2\vert V_\mathrm{ud}\vert^2}{8\pi^2}\left\{\frac{\nu f_\pi^2}{E_\nu^2 Q^2}\tilde L_{00}\right\}\frac{\sigma_\mathrm{max}}{4}\frac{M_R^2\Gamma^2}{(W^2-M_R^2)^2+M_R^2\Gamma^2}\left(1+3\cos^2\theta^\ast\right)\label{eq11}
\end{align}
with $\sigma_\mathrm{max} = 199\cdot10^{-27}\,$cm$^2$ and the other quantities referring to the resonance. This equation is analogous to equation (\ref{eq4}) with the difference that now the variables are determined by two body kinematics. The process does not exhibit a diffractive peak. For the dependence on the t variable we must substitute in equation (\ref{eq11})
\begin{align}
	\cos\theta^\ast=1+\frac{t}{2{p^\ast}^2}.\label{eq12}
\end{align}
The terms from the vector current squared and the interference have the same angular dependence so that the complete contribution from the $\Delta$-resonance can be analyzed in the manner described. This way we obtain the $t$ dependence for the background. Another criterion for separating the coherent from the background is provided by the opening angle of the pion relative to the neutrino direction. Starting with the triple differential cross section in equation (\ref{eq11}), we transformed it to the laboratory variables by computing the appropriate Jacobian. The results for a $^{12}$C target with an averaging over protons and neutrons are shown in figure (\ref{fig4}). We selected several values for $E_\nu$, $E_\pi=0.4\,$GeV and set the azimuthal angle $\phi$ equal to zero. This figure can be compared with the plots for coherent scattering in figure (2) of ref. [21] and figure (\ref{fig5}) in this article. The incoherent cross section is relatively large and extends to larger values of $\theta_\pi$. It peaks at $\theta_\pi\approx 40\,^\circ$ in contrast to coherent scattering which peaks at $\theta_\pi\approx 10\,^\circ - 15\,^\circ$ and vanishes for $\theta_\pi\approx 30\,^\circ$. A pure coherent signal appears only for $\theta_\pi< 15^\circ$.


The PCAC approach we described is valid for small values of $Q^2$. When we compare the theoretical curves with experimental points we observe that for $E_\nu < 4\,$GeV the four curves are close together. Thus small values of $Q^2_\mathrm{max}$ are acceptable for the comparison with the data. For energies $E_\nu > 4\,$GeV the experimental points agree with the theoretical curves where $Q^2_\mathrm{max}$ is closer to $1.0\,$GeV. This may be a consequence of the fact that for $\nu>1.0\,$GeV we used constant values for $a(\nu)$ and $b(\nu)$. We did this because we could not find data at these energies. If we select smaller values for $b(\nu)$ then the coherent cross section will be larger also for smaller values of $Q^2$. This observation suggest that we should try a fit by restricting $Q^2_\mathrm{max} = 0.2$ or $0.3$\,GeV$^2$ and then searching for values of $a(\nu)$ and $b(\nu)$ which reproduce the coherent data .

\vspace{.8cm}\noindent\textbf{\Large{4. SUMMARY}}\vspace{.8cm}\\
In the first chapter of the article we explained how chiral symmetry is applied to neutrino and antineutrino reactions. It was emphasized that the dominant contribution does not come from the pion pole, but from the cloud of mesons that surround the target [3,22]. In fact for low values of the neutrino energy the relevant range of $Q^2$ is small, so that the introduction of a form factor or of a propagator from axial mesons is not necessary. The new data indicate that for $E_\nu\gtrsim5\,$GeV, higher values of $Q^2_\mathrm{max}$ may be necessary. For comparisons with new data we took the theoretical curves from earlier calculations [4] and [5] and used the same curves for neutrinos and antineutrinos. For coherent scattering the integrated cross sections are expected to be equal and the data cofirm this expectation. In general, the agreement between theory and experiment is good, but the statistical uncertainties are still large.

Only at the first point with $E_\nu \approx E_{\bar\nu} \approx 1.75$\,GeV there is a difference between neutrinos and antineutrions We point out that the background at this energy is not diffractive and special care is necessary in order to determine the background from the $\Delta$-resonance which has a different dependence on the variable $t$.
It may also be helpful to define variables relative to the neutrino direction. In the econd chapter we summarized kinematic relations written in terms of variables relative to the neutrino direction. In figure (\ref{fig4}) we show the angular distributions of pions relative to the neutrino direction originating from the production of the $\Delta$-resonance. We computed them for several neutrino energies and for $^{12}$C target by taking the sum of scatterings on protons plus neutrons and then multiplied the sum by six. The angular deependence of the baclground is broader than the distribution for coherent scattering shown in figure 5. These distributions were computed for $Q^2 < 0.20\,$GeV$^2$. A recent calculation [23] presented angular distributions without any restrictions on the range of $Q^2$. We hope that new estimates will be helpful for understanding the background and improving the accuracy of coherent scattering.

\vspace{.8cm}\noindent\textbf{\Large{4. ACKNOWLEDGEMENT}}\vspace{.8cm}\\
This is an expanded version of a talk presented at the CETUP-Workshop in the summer of 2014. One of us (Emmanuel A. Paschos) wishes to thank the organizers of the workshop for providing a stimulating atmosphere that lead to the improvements discussed in this article.

\newpage
\vspace{.8cm}\noindent\textbf{\Large{REFERENCES}}\vspace{.8cm}\\
\begin{itemize}
\setlength{\itemsep}{-2pt}
\setlength{\parskip}{0pt}
\setlength{\parsep}{0pt} 
	\item[\text{[1]}]\ \ 	S. L. Adler, Phys. Rev. 135, 963 (1964) \\
	\item[\text{[2]}]\ \ 	D. Rein and L. Sehgal, Nucl. Phys. B223, 29 (1983) \\
	\item[\text{[3]}]\ \	A. Kartavtsev, G. Gounaris, E. A. Paschos, Phys. Rev. D74, 054007 (2006) \\
	\item[\text{[4]}]\ \	E. A. Paschos and D. Schalla, Phys. Rev. D80, 033005 (2009) \\
	\item[\text{[5]}]\ \	Ch. Berger and L. Sehgal, Phys. Rev. D79, 053003 (2009) \\
	\item[\text{[6]}]\ \	L. Alvarez-Ruso et al., Phys. Rev. C75, 055501 (2007) and Phys. Rev. C76, 068501 (2007) \\
	\item[\text{[7]}]\ \	S. K. Nakamura et al., Phys. Rev C81, 035502 (2010)\\
	\item[\text{[8]}]\ \	S. K. Singh et al., Phys. Rev. Lett. 96, 241801 (2006) \\
	\item[\text{[9]}]\ \	L. E. Amaro et al., Phys. Rev. D79, 013002 (2009) \\
	\item[\text{[10]}]\ \	E. Hernandez, Phys. Rev. D80, 013003 (2009) \\
	\item[\text{[11]}]\ \	T. Leitner, U. Mosel and S. Winkelmann, Phys. Rev. C79, 057601 (2009) \\
	\item[\text{[12]}]\ \	J.G. Morfin, J. Nieves and J. T.Sobczyk Adv. High Energy Phys. 93497 (2012) \\
	\item[\text{[13]}]\ \	R. M. Edelstein et al., Phys. Rev. 122 (1961) p.252 \\
	\item[\text{[14]}]\ \	F. G. Binon et al., Nucl. Phys. B17, 168 (1970) \\
	\item[\text{[15]}]\ \	T. Takahashi et al, Phys. Rev. C51, 2542 81995 \\
	\item[\text{[16]}]\ \	A. Higuera et al., arXiv, 1409.3853 (2014) \\
	\item[\text{[17]}]\ \	M. Hasegawa et al., Phys. Rev. Lett. 95, 252301 (2005) \\
	\item[\text{[18]}]\ \	K. Hiraide et al., Phys. Rev. D78, 112004 (2008) \\
	\item[\text{[19]}]\ \	H. J. Grabosch et al., Z. Phys. C31, 203 (1986) \\
	\item[\text{[20]}]\ \	R. Acciari et al., arXiv 1408.0598 (2014) \\
	\item[\text{[21]}]\ \	A. Higuera and E. A. Paschos, Eur. Phys. J. Plus 129 (2014) \\
	\item[\text{[22]}]\ \	B. Z. Kopeliovich et al., EPJ Web Conf. 70,0012 (2014) and arXiv [hep-ph] 1210.0411 \\
	\item[\text{[23]}]\ \	I. Schienbein, et al., DO-TH 14-28, arXiv 1411.6637 [hep-ph] \\
\end{itemize}

\end{document}